\documentclass[fleqn,10pt]{wlscirep}
\usepackage[utf8]{inputenc}
\usepackage[T1]{fontenc}

\def  \bsig    {\mbox{\boldmath$\sigma$}}

\def    \v          {\mbox{$\widetilde{v}$}}

\title{Twist-Tunable Spin-to-Charge Conversion and Valley-Contrasting Effects in Graphene/TMDC Heterostructures}

\author[1,*]{I. Wojciechowska}
\author[1,$\dagger$]{A. Dyrdał}
\affil[1]{Faculty of Physics and Astronomy, Adam Mickiewicz University in Poznań, ul.Uniwersytetu Poznańskiego 2, 61-614 Pozna\'n, Poland}

\affil[*]{izabella.wojciechowska@amu.edu.pl}
\affil[$\dagger$]{adyrdal@amu.edu.pl}

\begin{abstract}
We  consider graphene deposited on monolayers of such transition-metal dichalcogenides like MoSe$_2$, WSe$_2$, MoS$_2$, and WS$_2$. Our key objective is to study the impact of relative twist angle between the monolayers on the proximity-induced spin-orbital  effects and orbital phenomena in graphene. To do this we used an effective model Hamiltonian for low-energy states, taken from available literature.  The Green function formalism is used to  calculate analytical formula for  the spin Hall effect and nonequilibrium spin polarization in the system. 
We also determine the valley Hall and valley polarization  effects, and their dependence on the twist angle.  We have shown that the valley Hall conductivity can take the quantized value equal to $\pm 2 e^2/h$.
\end{abstract}
\begin{document}

\flushbottom
\maketitle
\thispagestyle{empty}

%\noindent Please note: Abbreviations should be introduced at the first mention in the main text – no abbreviations lists. Suggested structure of main text (not enforced) is provided below.

\section{Introduction}

Spin-orbit coupling (SOC) plays an essential role in the spin-dependent electronic properties of low-dimensional materials and spintronics applications. Although intrinsic SOC is weak in many individual two-dimensional (2D) crystals, it can be significantly enhanced through the so called proximity effects.
Proximity-induced spin-orbit coupling arises when a 2D material with negligible intrinsic SOC, such as graphene, is placed in contact with a material possessing strong SOC, e.g., transition metal dichalcogenides (TMDCs) or topological insulators (TIs)~\cite{wang_strong_2015,PhysRevB.92.155403,avsar_spinorbit_2014,Mendes2014,cao_unconventional_2018,Sierra2021,Kou2014,Jin2013,Song2018,Yang2013}. The nature and magnitude of the induced SOC depend on the interlayer hybridization, symmetry, and stacking configuration. 

The rapid advancement in the fabrication of van der Waals (vdW) heterostructures~\cite{Geim2013,Liu2016,Ajayan2016}—combined with the discovery of correlated phases and superconductivity in magic-angle twisted bilayer graphene~\cite{Hennighausen_2021,cao_unconventional_2018,Shen2022}—gave rise to the emerging field of spintronics that is called twistronics~\cite{Carr2020}.
Recent theoretical and experimental studies have already demonstrated that the twist angle between layers in vdW heterostructures strongly modifies interfacial interactions and allows for the control of band structure and symmetry-breaking effects~\cite{Leutenantsmeyer_2DMaterials2017,zollner_twist-_2023,Zihlmann_PhysRevB.97.075434,Gonçalves_2022}. In particular, rotational misalignment leads to the formation of moiré superlattices, which generate spatially varying interlayer hybridization and result in band reconstruction, mini-band formation, and altered Berry curvature distributions~\cite{cao_unconventional_2018,Chebrolu_PhysRevB2019,Efimkin_PhysRevB2018}.  These moiré-induced effects directly influence the spin-orbit texture and can lift spin and valley degeneracies in a manner that is strongly dependent on the twist angle. As such, the twist angle serves as a critical tuning parameter for engineering proximity-induced SOC, potentially enabling phase transitions between distinct topological regimes~\cite{Xian_NatCommun2021,Jin_NatMat2021,Naimer_PRB2024}.

%Recent first principle studies on twisted graphene on TMDCs indicate strong twist-angle dependent 

In this paper we consider the effect of the relative twist between graphene and  monolayer of semiconducting TMDC monolayer on the certain transport characteristics of such heterostructure. Specifically, we consider t-Gr/MoSe$_2$, t-GrWSe$_2$, t-Gr/MoS$_2$, and t-Gr/WS$_2$ (t-Gr abbreviates twisted graphene). The elecronic band structures of the four above heterostructures have been considered recently based on density functional theory (DFT)~\cite{li_twist-angle_2019,zollner_twist-_2023,Naimer_PRB2024} and the parameters describing effective low-energy Hamiltonian of graphene fitted in these works allows us to investigate selected transport characteristics using linear response theory and Green function formalism that lead us to the fully analytical final results. The paper is organised as follows. In Section \ref{sec:Sec2} we present the low energy effective  Hamiltonian describing electronic properties of twisted graphene deposited on TMDC monolayer and twist-angle dependence of effective parameters defining Hamiltonian. We show also the electronic band structure of graphene deposited on certain semiconducting TMDC monolayer and how the twist angle change its shape as well as spin-momentum locking.
In Section \ref{sec:Sec3} and \ref{sec:Sec4}  we present results of our theoretical study of spin-to-charge conversion and valley contrasting effects using Green function formalism and linear response theory. In Sec. \ref{sec:Sec3} it is shown how the twist angle affects the spin-to-charge conversion phenomena, i.e. spin Hall effect and Rashba Edelstein effect. In Sec. \ref{sec:Sec4} the valley Hall effect and valley nonequilibrium spin polarization are presented and discussed.
The discussion and summary of our results are presented in Sec.~\ref{sec:Sect5}.

\begin{figure}[t]
\includegraphics[width=0.95\columnwidth]{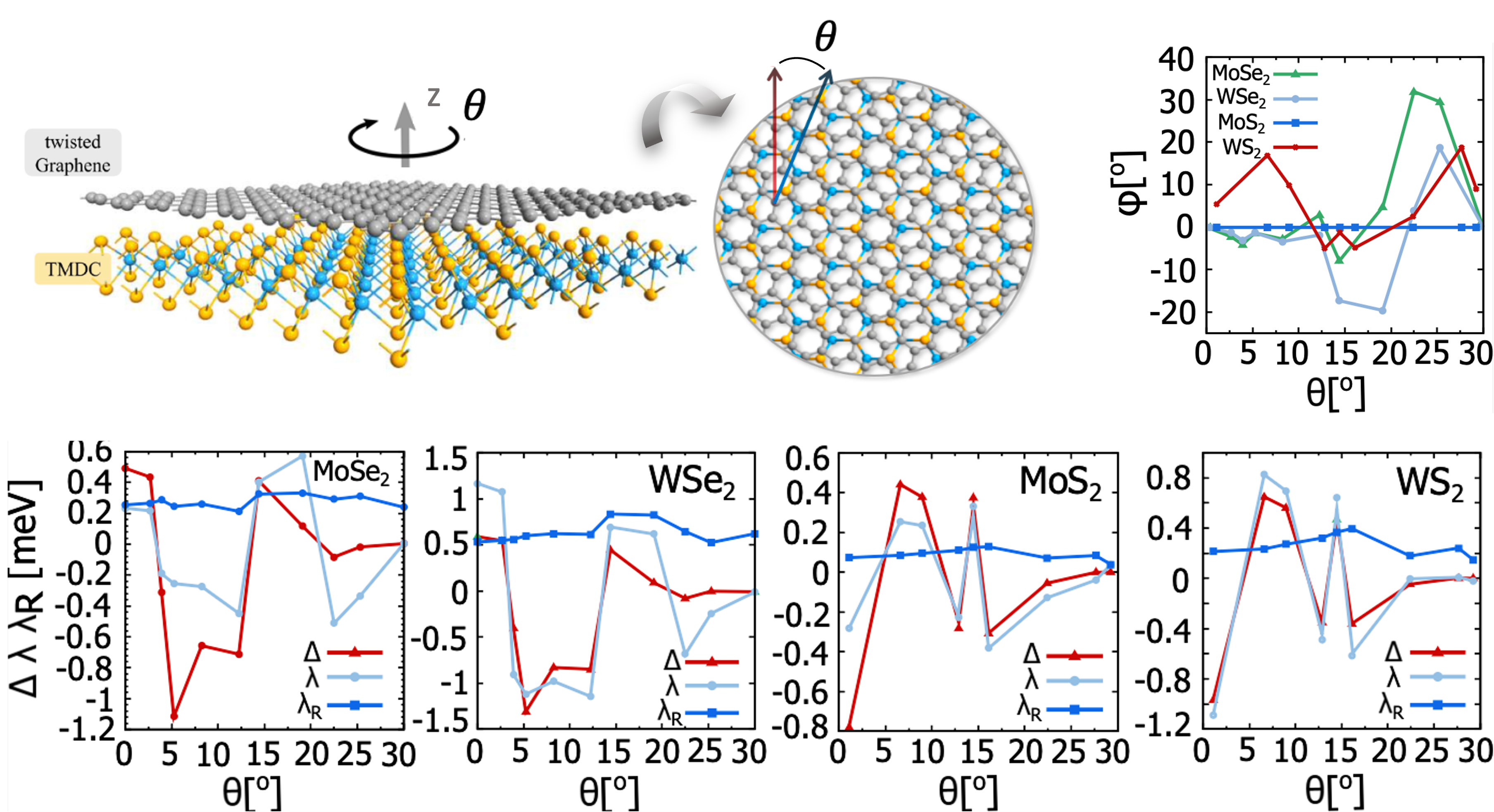}
\caption{Schematic picture of graphene twisted by the angle $\theta$ with respect to the monolayer of TMDC (side and top view) and parameters defining Hamiltonian (\ref{eq:Ham_gen}) as a function of twisted angle $\theta$ for four semiconducting transition metal dichalcogenides: MoSe$_2$, WSe$_2$, MoS$_2$, and WS$_2$. Data are taken from Ref.(\citeonline{zollner_twist-_2023}).}
\label{fig:Fig1}
\end{figure}

\section{Model}
\label{sec:Sec2}

We consider graphene deposited on a monolayer of one of the semiconducting  transition metal dichalcogenices, like MoSe$_{2}$, WSe$_{2}$, MoS$_{2}$ and WSe$_{2}$.   The effective Hamiltonian describing low-energy electronics states around the K/K' points of the Brillouine zone of graphene in proximity to TMDCs takes the following form~\cite{david_induced_2019,PhysRevB.99.075438,zollner_twist-_2023,Gmitra_Fabian_PRB2015}:
\begin{equation}\label{eq:Ham_gen}
		 \hat{H}^{\nu} = {\hat{H}}_{\scriptscriptstyle{0}}^{\nu}+{\hat{H}}_{\scriptscriptstyle{\Delta}}+{\hat{H}}_{\scriptscriptstyle{I}}^{\nu}+{\hat{H}}_{\scriptscriptstyle{R}}^{\nu}\, ,
\end{equation}
where  the individual terms of the above Hamiltonian read:
\begin{align}
\hat{H}_{\scriptscriptstyle{0}}^{\nu} =& \upsilon (\nu k_x \sigma_x - k_y \sigma_y)\otimes s_0,\\
\hat{H}_{\scriptscriptstyle{\Delta}} =& \Delta \sigma_z \otimes s_0,\\ \label{eq:Ham_lamdba}
\hat{H}_{\scriptscriptstyle{I}}^{\nu} =& \nu (\lambda_{\textrm{I}}^\textrm{A} \sigma_{+}+\lambda_{\textrm{I}}^\textrm{B} \sigma_{-})\otimes s_z,\\ \label{eq:Hami_rashba}
\hat{H}_{\scriptscriptstyle{R}}^{\nu} =& -\lambda_{\textrm{R}}\textrm{e}^{-\textrm{i}\varphi\frac{s_z}{2}}(\nu \sigma_x \otimes s_y + \sigma_y \otimes s_x)\textrm{e}^{\textrm{i}\varphi\frac{s_z}{2}}. 
\end{align}
We have used above the following notation: $\nu = \pm1$ selects K or K' valley, respectively,
 $k_{x,y}$ are the components of the wavevector, i.e., $\mathbf{k} = (k_x, k_y)$ and $k_{x}^{2} + k_{y}^{2} = k^2$; matrices $\hat{\sigma}_{0}, \hat{\bsig} = (\sigma_{x}, \sigma_{y}, \sigma_{z})$ denote the identity matrix and Pauli matrices acting in the pseudospin space, whereas $\hat{s}_{0}, \hat{\mathbf{s}} = (s_{x}, s_{y}, s_{z})$ define identity and Pauli matrices acting in the spin space.
The first term of the Hamiltonian~(\ref{eq:Ham_gen}), captures the orbital physics of pristine graphene ($v= \hbar v_{\textrm{F}}$ with  $v_{\textrm{F}}$ denoting Fermi velocity). The second term of ~(\ref{eq:Ham_gen}), $\hat{H}_{\Delta}$, describes the staggered potential arising due to sublattice symmetry breaking.
The last two terms in Eq.~(\ref{eq:Ham_gen}) describe two possible components of the spin-orbit coupling, that may appear in the structure: the so-called intrinsic spin-orbit interaction and Rashba coupling. The intrinsic SOC, given by Eq.~(\ref{eq:Ham_lamdba}), has a subblattice dependent form with the copuling parameters $\lambda_{I}^{A,B}$. The Rashba spin-orbit interaction is given by the generalised form that containing two parameters: the coupling amplitude, $\lambda_{R}$, and the Rashba angle, $\phi$, determining the spin-momentum locking in the system.

\begin{figure}[t]
\includegraphics[width=0.999\columnwidth]{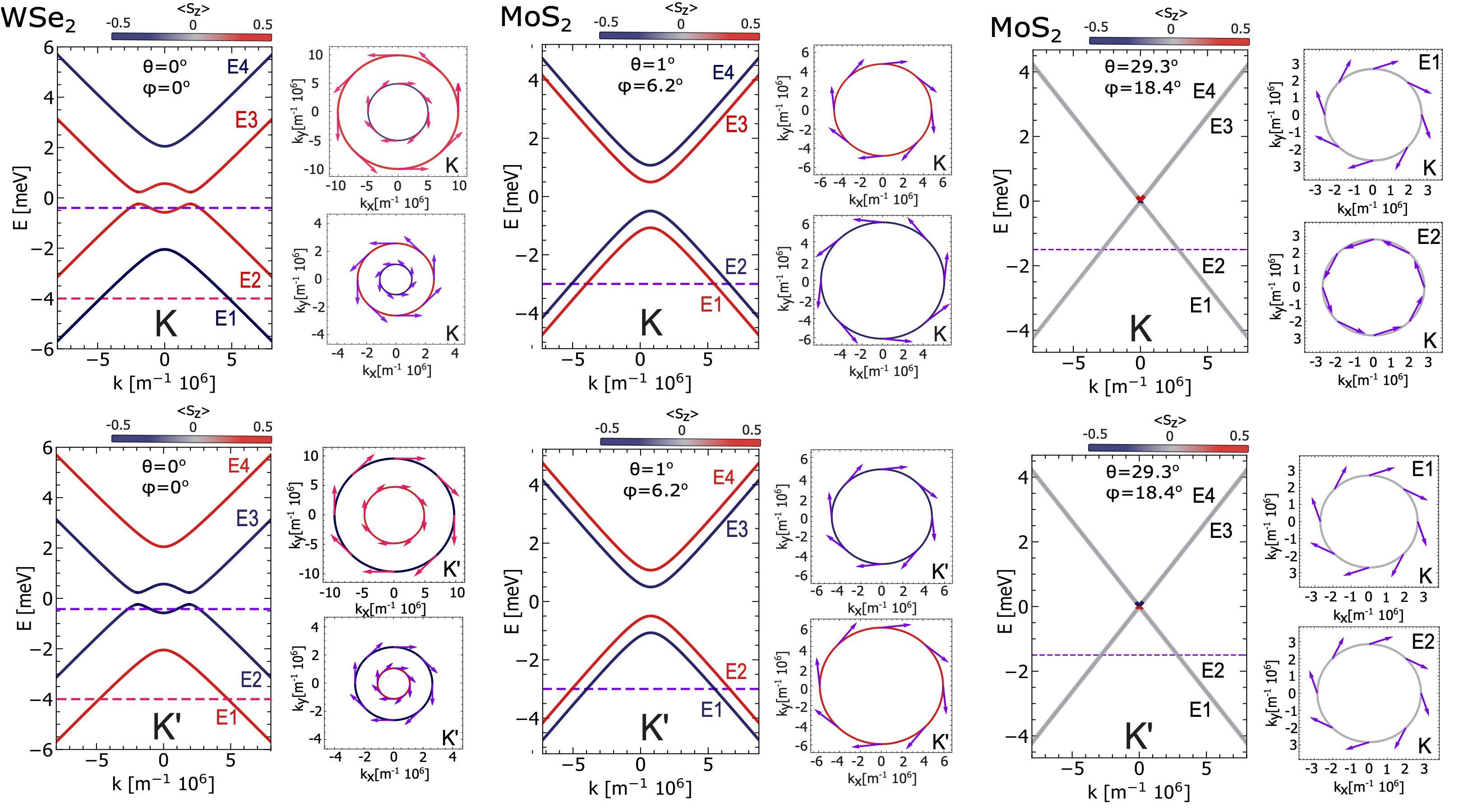}
\caption{The energy dispersions plotted for t-Gr/WSe$_2$ for the twist angle $\theta = 0 \deg$ and the Rashba angle $\phi = 0 \deg$ and for t-Gr/MoS$_2$ for $\theta = 1 \deg$ and the Rashba angle $\phi = 6.2 \deg$ as well as for $\theta = 29.3 \deg$ and the Rashba angle $\phi = 18.4 \deg$. The color of the band lines
corresponds to the $s_z$ spin expectation value whereas the in-plain spin expectation values have been indicated on energy contours. The values of parameters $\lambda$, $\lambda_{R}$, $\phi_R$ and $\Delta$ for the certain twist angle, $\theta$, are taken from Fig~\ref{fig:Fig1}, the $\upsilon$ parameter is equal $5.414\cdot10^{-10}$ eVm for  WSe$_2$ and $4.356\cdot10^{-10}$ eVm for MoS$_2$. }
\label{fig:Fig2}
\end{figure}

The eigenvalues of Hamiltonian~(\ref{eq:Ham_gen}) take the following form
\begin{align}
\label{eq:simpE_14}
		E_{1,4} =  \mp \sqrt{k^2v^2 + \Delta^2 + \lambda^2 + 2\lambda_{\scriptscriptstyle{R}}^2  + 2\xi},\\
\label{eq:simpE_23}
		E_{2,3} =  \mp \sqrt{k^2v^2 + \Delta^2 + \lambda^2 + 2\lambda_{\scriptscriptstyle{R}}^2  -  2\xi},
\end{align}
where $\xi = \sqrt{(\Delta\lambda + \lambda_{\scriptscriptstyle{R}}^2)^2 + k^2v^2(\lambda^2 + \lambda_{\scriptscriptstyle{R}}^2)^2}$. Importantly, all the parameters defining the eigenvalues $E_{1-4}$ depend on material parameters and are also strongly dependent on the twist angle $\theta$, as presented in Fig.~\ref{fig:Fig1}.  In Fig.~\ref{fig:Fig2} the low-energy electronic states of graphene deposited on a single-layer of TMDCs are presented for fixed values of the twist angle. The expectation values of the spin at certain wavevectors are also presented there. The selected examples reflect the possible behaviour of the eigenstates, that one can obtain for a fixed twist angle. Accordingly, the low-energy bands of t-Gr/WSe$_2$ for the twist angle $\theta = 0\deg$ and Rashba angle $\phi = 0 \deg$ are strongly spin-splitted with a characteristic 'Mexican hat' shape of the top-most valence and bottom-most conduction bands, which is typical for the case when the staggered potential, $\Delta$, dominates over the intrinsic spin-orbit coupling strength, $\lambda$.  In turn, for t-Gr/MoS$_2$ structure with the $\theta$ equal  1$\deg$ and $29.3\deg$ one finds $|\lambda| < |\Delta|$ and the spin splitting occurs vertically (the energy bands have only single extremal point at $k = 0$). Note that the vanishing of staggered potential and intrinsic SOC leads to the almost degenerated bands and the vanishing of the bandgap (this is the case of t-Gr/MoS$_2$ for the twist angle $\theta = 29.3$ deg).

\section{Twist-angle tuneable spin-to-charge conversion}
\label{sec:Sec3}

The intentional twist of graphene with respect to TMDC allows significant tuning of spin proximity effects and, consequently, the tuneability of the spin-to-charge conversion effects. 
In this section we discuss the behaviour of the two most important effects for spin-orbitronics. i.e., spin Hall effect and current-induced spin polarization, also known as Rashba-Edelstein effect (REE) or inverse spin galvanic effect~\cite{Dyakonov1971,Hirsch1999,Aronov1991,Edelstein1990Jan,Golub2011Sep,Ganichev2001May,Ganichev2002May,Sinova_RevModPhys2015}. Using the linear response theory and Green function formalism one can define the spin Hall conductivity, SHC, in the considered system as follows:
\begin{equation}
\sigma_{SH} = \sigma_{xy}^{s_z\,K} + \sigma_{xy}^{s_z\,K'}.
\end{equation}
Thus, the total spin Hall conductivity is a sum of the transverse spin Hall conductivity (calculated in the dc limit) for the two inequivalent valleys, that is for $\tau = K,K'$, according to the formula~\cite{Abrikosov,Mahan,Dyrdal2017,wojciechowska_intrinsic_2024}:
\begin{equation}
\label{eq:sigma_spin_yx}
\sigma_{yx}^{s_{z}\,\tau} = \lim_{\omega \to 0} \frac{e^{2} \hbar}{\omega} \int \frac{d \varepsilon}{2\pi} \int \frac{d^{2}\mathbf{k}}{(2\pi)^{2}} \mathrm{Tr} \left[\hat{j}_{y}^{s_z\,\tau} G^{\tau}_{\mathbf{k}}(\varepsilon + \omega) \hat{\upsilon}_{x}^{\nu}G^{\tau}_{\mathbf{k}}(\varepsilon ) \right]
\end{equation}
where the spin current density operator is defined as $\hat{j}_{y}^{s_z\,\nu} = \frac{1}{2}[\hat{\upsilon}_{y}^{\nu},\hat{s}_{z}]_{+}$ with the velocity, $\hat{\upsilon}_{\alpha}^{\nu}$, and spin, $\hat{s}_{z}$, operators defined as $\hat{\upsilon}_{\alpha} = \frac{1}{\hbar} \frac{\partial \hat{H}^{\nu}}{\partial k_{\alpha}}$ ($\alpha = {x,y}$), $\hat{s}_{z} = \frac{\hbar}{2}\hat{I}_{2x2} \otimes \hat{s}_z$, respectively. $G_{\mathbf{k}}^{\nu}(\varepsilon)$ is the casual Green's function defined as $G_{\mathbf{k}}^{\nu}(\varepsilon) = [(\varepsilon + \mu + i\delta \mathrm{sign}(\varepsilon))\sigma_{0} \otimes \hat{s}_{0} - \hat{H}^{\nu}]^{-1}$  where $\mu$ denotes the chemical potential, and $\delta \to 0^{+}$ (as we consider the clean limit).

Taking into account the two possible cases corresponding to the two possible behaviours  of the spin-splitted bands (i.e. 'Mexican hat' or 'vertical like'), we found analytical expressions describing the spin Hall conductivity in the all energy regions. In general, in our system the formulas for the spin Hall  conductivity can be defined in seven distinct energy ranges.
Accordingly, when the bands are spin-splitted in a vertical-like way, that is when $0 <|\lambda| < |\Delta|$, one finds the following expressions valid for the specific ranges of the Fermi energy. Thus, when Fermi energy crosses both conduction or both valence bands, i.e., when $\left|\mu\right| > \sqrt{(\Delta +\lambda )^2+4\lambda_{\scriptscriptstyle{R}}^2}$ the spin Hall conductivity takes the form
\begin{equation}
\label{eq:expr1}
\tiny{
		\sigma_{SH} = \mp\frac{\Delta  \lambda ^3 }{2 \left(\lambda ^2+\lambda_{\scriptscriptstyle{R}}^2\right)} \left(\frac{1}{\zeta_{\scriptscriptstyle{+}}}-\frac{1}{\zeta_{\scriptscriptstyle{-}}}\right) \pm \frac{\lambda_{\scriptscriptstyle{R}}^2}{4 \left(\lambda ^2+\lambda_{\scriptscriptstyle{R}}^2\right)^2}\left[ \lambda ^2\left(\frac{\chi_{\scriptscriptstyle{1+}}}{\zeta_{\scriptscriptstyle{+}}} - \frac{\chi_{\scriptscriptstyle{1-}}}{\zeta_{\scriptscriptstyle{-}}} \right) + \lambda_{\scriptscriptstyle{R}}^2\left(\frac{\chi_{\scriptscriptstyle{2+}}}{\zeta_{\scriptscriptstyle{+}}} - \frac{\chi_{\scriptscriptstyle{2-}}}{\zeta_{\scriptscriptstyle{-}}} \right) \right],
  }
\end{equation}
where we used the following notation: 
\begin{align}
   &\chi_{\scriptscriptstyle{1\pm}}=\pm 2\eta + \mu^{2}-2\lambda^{2}+(\lambda-\Delta)^{2}, \\
   &\chi_{\scriptscriptstyle{2\pm}}=\pm 2\eta + \mu^{2}-2\lambda^{2}-(\lambda-\Delta)^{2},\\
   &\zeta_{\scriptscriptstyle{\pm}}=\sqrt{\kappa \pm 2\eta(\lambda^{2}+\lambda_{R}^{2})},\\
   &\eta=\sqrt{\mu^{2}(\lambda^{2}+\lambda_{R}^{2})-(\lambda-\Delta)^{2}\lambda_{R}^{2}},\\
   &\kappa=(\Delta\lambda+ \lambda_{R}^{2})^{2} + (\mu^{2}+\lambda^{2}-\Delta^{2})(\lambda^{2}+\lambda_{R}^{2}).
\end{align}
When Fermi energy crosses only one valence or one conduction band, $\left| \lambda -\Delta \right| < \left|\mu \right| \le  \sqrt{(\Delta +\lambda )^2+4\lambda_{\scriptscriptstyle{R}}^2}$, one finds
\begin{eqnarray}
\label{eq:cond2}
   \sigma_{SH} = \pm \frac{\Delta  \lambda ^3 }{2 \zeta_{\scriptscriptstyle{+}} \left(\lambda ^2+\lambda_{\scriptscriptstyle{R}}^2\right)}
	\mp \frac{\lambda_{\scriptscriptstyle{R}}^2}{4 \zeta_{\scriptscriptstyle{+}} \left(\lambda ^2+\lambda_{\scriptscriptstyle{R}}^2\right)^2}\left( \lambda^2 \chi_{\scriptscriptstyle{1+}} + \lambda_{\scriptscriptstyle{R}}^2 \chi_{\scriptscriptstyle{2-}}\right)\hspace{4cm} \nonumber\\ 
    \hspace{2cm}\pm \frac{\lambda}{2\left(\lambda ^2+\lambda_{\scriptscriptstyle{R}}^2\right)^2 \left|\Delta  \lambda +\lambda_{\scriptscriptstyle{R}}^2\right|}\bigg[\lambda \lambda_{\scriptscriptstyle{R}}^2\left(\Delta^2 - \lambda\Delta - \lambda^2 \right) - \Delta\lambda^4 + \lambda_{\scriptscriptstyle{R}}^4\left(\Delta - 2 \lambda \right) \bigg].
\end{eqnarray}
For Fermi energy located between the top-most valence or bottom-most conduction band minima and maxima, i.e., \newline ${-\left|\lambda-\Delta\right| \sqrt{\frac{\lambda_{R}^{2}}{\lambda_{R}^{2}+\lambda^{2}}} \leq\mu  \leq \left|\lambda-\Delta\right| \sqrt{\frac{\lambda_{R}^{2}}{\lambda_{R}^{2}+\lambda^{2}}} }$  (this is when the bands have Mexican hat shape) one gets

\begin{equation}
\label{eq:expr23}
\tiny{
		\sigma_{SH} = \pm\frac{\Delta  \lambda ^3 }{2 \left(\lambda ^2+\lambda_{\scriptscriptstyle{R}}^2\right)} \left(\frac{1}{\zeta_{\scriptscriptstyle{+}}}-\frac{1}{\zeta_{\scriptscriptstyle{-}}}\right) \mp \frac{\lambda_{\scriptscriptstyle{R}}^2}{4 \left(\lambda ^2+\lambda_{\scriptscriptstyle{R}}^2\right)^2}\left[ \lambda ^2\left(\frac{\chi_{\scriptscriptstyle{1+}}}{\zeta_{\scriptscriptstyle{+}}} - \frac{\chi_{\scriptscriptstyle{1-}}}{\zeta_{\scriptscriptstyle{-}}} \right) + \lambda_{\scriptscriptstyle{R}}^2\left(\frac{\chi_{\scriptscriptstyle{2+}}}{\zeta_{\scriptscriptstyle{+}}} - \frac{\chi_{\scriptscriptstyle{2-}}}{\zeta_{\scriptscriptstyle{-}}} \right) \right].
  }
\end{equation}
Finally, 
for  $ -\left|\lambda-\Delta\right| \sqrt{\frac{\lambda_{R}^{2}}{\lambda_{R}^{2}+\lambda^{2}}} <\mu <\left|\lambda-\Delta\right| \sqrt{\frac{\lambda_{R}^{2}}{\lambda_{R}^{2}+\lambda^{2}}} $ (that is for the Fermi energy located inside the energy gap) the spin Hall conductivity vanishes, $\sigma_{SH} = 0$. Here, it should be also stressed that, according to the data obtained based on  DFT calculations, the above formulas have been obtained assuming that $\Delta \lambda > 0$. Moreover, when the top-most valence and bottom-most conduction bands change their shapes from the Mexican-hat-like dispersion to the dispersion with only a single extremal point, the energy range between  ${-\left|\lambda-\Delta\right| \sqrt{\frac{\lambda_{R}^{2}}{\lambda_{R}^{2}+\lambda^{2}}} \leq \varepsilon  \leq \left|\lambda-\Delta\right| \sqrt{\frac{\lambda_{R}^{2}}{\lambda_{R}^{2}+\lambda^{2}}} }$ vanishes and the energy gap is defined in the energy range between $-|\lambda - \Delta| < \mu < |\lambda - \Delta|$. This happens when $\lambda \ll \lambda_{R}$.

 \begin{figure}[t!]
\includegraphics[width=0.7\columnwidth]{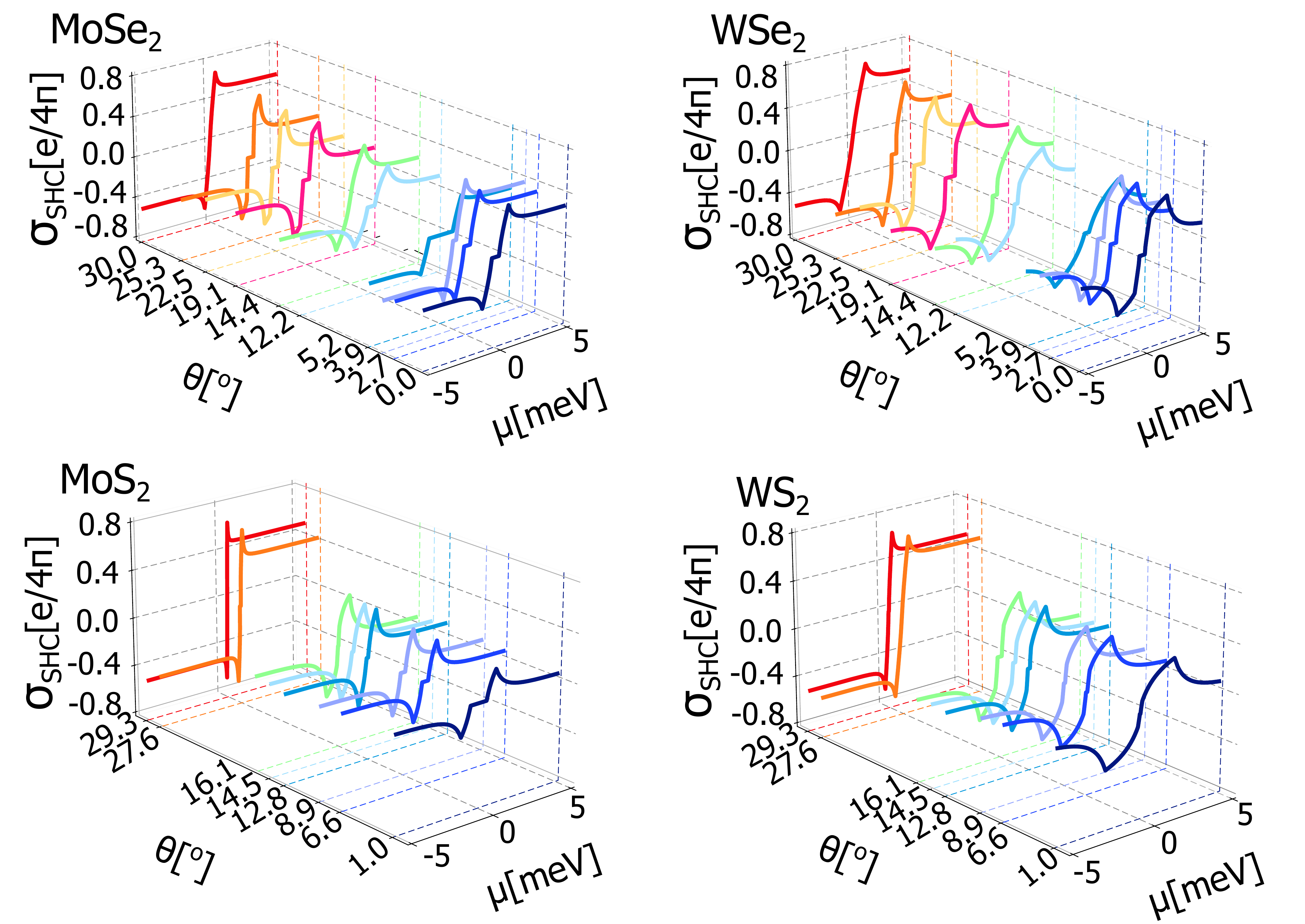}
\centering
	\caption {The spin Hall conductivity of graphene deposited on semiconducting TMDC monolayer ( i.e., t-Gr/MoSe$_{2}$, t-Gr/WSe$_{2}$, t-Gr/MoS$_{2}$ and t-Gr/WSe$_{2}$) as a function of the Fermi energy, $\mu$, for certain twist angle between graphene and TMDC, $\theta$. The values of parameters $\lambda$, $\lambda_{R}$, $\phi_R$ and $\Delta$ for the certain twist angle, $\theta$, are taken from Fig~\ref{fig:Fig1}, the $\upsilon$ parameter is equal $5.414\cdot10^{-10}$ eVm for  WSe$_2$, $4.348\cdot10^{-10}$ eVm for WS$_2$, $5.413\cdot10^{-10}$ eVm for MoSe$_2$, and   $4.356\cdot10^{-10}$ eVm for MoS$_2$.}
	\label{fig:Fig3}
\end{figure}

Figure \ref{fig:Fig3} shows the spin Hall conductivity plotted as a function of the Fermi energy, $\mu$,  for specific values of the twist angle,$\theta$, in the range between $0\deg$ and $30\deg$, and for   t-Gr/MoSe$_{2}$, t-Gr/WSe$_{2}$, t-Gr/MoS$_{2}$ and t-Gr/WSe$_{2}$. One can easily note that the twist of graphene layer with respect to the TMDC monolayer can significantly modulate the spin Hall conductivity.  We also note that the spin Hall  conductivity behaves antisymmetric with respect to the sign of Fermi energy, and reveals a sharp peak when Fermi energy crosses the minimum of the second conduction band and maximum of the bottom-most valence band. Such a behaviour is expected in proximitized graphene systems with Rashba SOC. However, the spin Hall conductivity does not depend on the Rashba angle $\phi$ (what clearly follows from the obtained analytical formulas).
%Upon the certain twist that change  varies for different angles achieving extreme values around $|0.8|$ e/4$\pi$ for all structures. 
It should be also stressed, that in the considered heterostructures, the intrinsic SOC, $\lambda$, is of the valley-Zeeman type, thus we do not observe the spin Hall response when Fermi energy is in the energy gap (there is no contribution from the Fermi see to the SHC). 
%what indicates absence of nontrivial topological phase transition.

%In this paper we analytically consider intrinsic contribution to different spin-orbit coupling driven phenomena. The nature of this contribution depends purely on band structure which enable detailed analysis. In our model we observe  spin-orbit coupling of two kinds: proximity induced SOC and Rashba SOC. DFT calculations have showed that parameters satisfy $\lambda_{\textrm{I}}^\textrm{A} \approx -\lambda_{\textrm{I}}^\textrm{B}$  what makes intrinsic SOC of valley Zeeman type. 

Based on the plots presented in Fig.\ref{fig:Fig1}, it is seen that the fitted Rashba SOC amplitude, $\lambda_{R}$, is rather not affected by the twist angle. In contrast, the staggered potential, $\Delta$, and intrinsic SOC, $\lambda$, can significantly change with the twist. Accordingly, the change of $\Delta$ and $\lambda$ with the twist angle controls the width of the energy gap and consequently sets a specific energy range where we do not observe any spin current. From the plots presented in Fig. \ref{fig:Fig3} one can note that SHC achieves its largest values for the twist angles, $\theta$, equal $30 \deg$(MoSe$_{2}$,WSe$_{2}$) and $29.3 \deg$ (MoSe$_{2}$,WSe$_{2}$). These angles correspond to the case when the parameter $\lambda$ tends to zero, which is caused by emerging mirror plane symmetry. Under these circumstances the Rashba SOC dominates in the system.  

%The intrinsic nature of examine contribution reflects in the shape of particular SHC curves.  Since each twist induce  symmetry breaking it affects the band structure.  We can distinguish  two cases of general character of dispertion relation  that can appear in this model (\ref{Character}) relaying on value of $\Delta$ and $\lambda$. Thus, there are also two main characters of SHC curves i.e. for $MoSe_{2}$: $30^{o}$() and $22.5^{o}$()

\begin{figure}[t!]
\includegraphics[width=0.75\columnwidth]{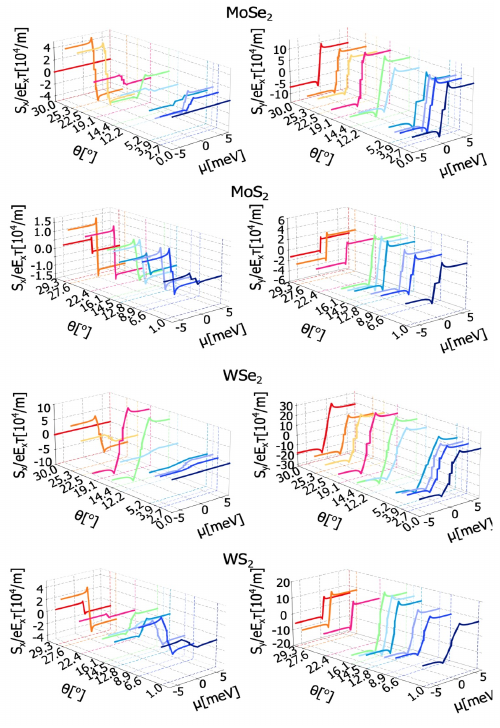}
\centering
\caption {The \textit{x} and \textit{y} components of nonequilibrium spin polarization plotted as a function of Fermi energy, $\mu$, for specific values of the twist angle, $\theta$. The values of parameters $\lambda$, $\lambda_{R}$, $\phi_R$ and $\Delta$ for the certain twist angle, $\theta$, are taken from Fig~\ref{fig:Fig1}, the $\upsilon$ parameter is equal $5.414\cdot10^{-10}$ eVm for  WSe$_2$, $4.348\cdot10^{-10}$ eVm for WS$_2$, $5.413\cdot10^{-10}$ eVm for MoSe$_2$, and   $4.356\cdot10^{-10}$ eVm for MoS$_2$. }
\label{fig:Fig4}
\end{figure}

The nonequilibrium spin polarization arising in the system as a consequence of the Rashba-Edelstein effect (REE) can be  calculated within the Green function formalism, too. Accordingly, the components of the nonequilibrium spin densities can be calculated starting from the expression:
\begin{equation}
S_{\alpha} = S_{\alpha}^{K}+S_{\alpha}^{K'}
\end{equation}
where~\cite{Abrikosov,Mahan,dyrdal_current-induced_2014,Dyrdal_PRB2015}
 \begin{equation}
 \label{eq:S-alpha-nu}
S_{\alpha}^{\nu} = \lim_{\omega \rightarrow 0} \frac{e E_{x} \hbar}{\omega} \int\frac{d^{2}\mathbf{k}}{(2\pi)^{2}} \int \frac{d \varepsilon}{2\pi} \mathrm{Tr} \left[\hat{s}_{\alpha} G^{\nu}_{\mathbf{k}}(\varepsilon + \omega) \hat{\upsilon}_{x}^{\nu}G^{\nu}_{\mathbf{k}}(\varepsilon) \right].
 \end{equation}
The above expression in the dc-limit takes the form:
\begin{equation}
\label{eq:S_alpha^tau(EF)}
		S_{\alpha}^{\nu} = \dfrac{e E_{x}\hbar}{2\pi}\int\,\frac{d^{2}\mathbf{k}}{(2\pi)^{2}}\mathrm{Tr} \left[\hat{s}_{\alpha}G_{\mathbf{k}}^{R\,\nu}\hat{\upsilon}_{x}^{\nu}G_{\mathbf{k}}^{A\,\nu}\right]
\end{equation}
that contains contribution from the states at the Fermi level. Also in this case we obtained fully analytical formulas, however they are rather cumbersome, so we decided to show only the numerical results, as presented in Fig.\ref{fig:Fig4}.  As one can note, the nonequilibrium spin polarization for each of the considered system contains two nonzero components, that are proportional to the external electric field, $E_x$ and relaxation time, $\tau$. Both nonzero components are   oriented in the plain of the structure, i.e., the $y$-component is oriented perpendicularly to the external electric field and the $x$-component is oriented parallel to the electric field. The component parallel to the external electric field, recently called the unconventional  Rashba Edelstein effect~\cite{zollner_twist-_2023,lee_charge--spin_2022}, is quite naturally expected in systems with more complicated forms of spin-orbit coupling~\cite{Schliemann_PRB2007,Schliemann_2019}. In the heterostructures considered in this paper, the nonzero $S_{x}$ component of spin polarization is a simple consequence of the nonzero Rashba angle, $\phi$, that governs the spin-momentum locking and deflects the expectation value of spin out of the perpendicular orientation to the quasiparticle momentum ($\phi = 0\deg$). As the maximum Rashba angle $|\phi|$, appearing for a certain twist angle, is about 20 deg for t-GR/WS$_2$ and t-Gr/WSe$_2$ and 30 deg for MoSe$_2$, thus the $x$ component of the spin polarization is always smaller than the $y$ component.

Fig.~\ref{fig:Fig4} also clearly shows that both $S_{x}$ and $S_{y}$ components behave anti-symmetrically  with respect of the sign reversal of the Fermi energy, and their magnitudes are strongly modulated by the twist angle. Importantly, the sign of the $S_{x}$ component can be changed for those twist angles for which the Rashba angle is negative.

%Rashba Edelstein effect know also as current induced spin polarization (CISP) is spin-orbit coupling driven phenomenon observed in systems with Rashba spin orbit coupling. In the case of twisted graphene/TMDC heterostructures analyzed in this paper, we observe not only a component lying within the plane of graphene and perpendicular to the applied electric field (Fig. \ref{CISP_Sy}), but also a component oriented in the same direction as the applied electric field (Fig. \ref{CISP_Sx}). It is caused by additional parameter intoduced by twist - Rashba angle $\varphi$ and because of that spin polarization in x direction is dependent and can be controlled only by that variable. In order to show it more precisely we focus on one example meaning $MoSe_{2}$ case in Fig. \ref{CISP_Sx} and $\varphi$ in  Fig. \ref{Parameters}. One can notice that the magnitude of systems response is dependent on value of Rashba angle what prooves previous statement. 

\section{Valley Hall effect and valley Rashba-Edelstein effect controlled by the twist angle of graphene}
\label{sec:Sec4}

Graphene based materials are currently a platform for the development of  valley-contrasting phenomena. The corresponding field of research, referred to as  valleytronics, is focused on an active use of the additional electron degree of freedom due to  valleys (local minimum/extremum in the electronic band structure). The graphene based heterostructures with the two inequivalent valleys at K and K' points of the Graphene's Brillouine zone provide excellent platform for exploring the valley physics and valley contrasting phenomena. The possibility of usage in data storage and logic devices another degree of freedom, in addition to the charge and spin ones,  is extremely promising and has focused a lot of attention in recent years by theory and experimental groups. %\cite{Manchon_PRB2016}.  
The valley Hall effect can be detected experimentally, for instance, with the use of the Kerr rotation microscopy~\cite{Mak_Science2014,Lee_Mak_NatNano2016}.

\begin{figure} [t!]
\centering
\includegraphics[width=0.85\textwidth]{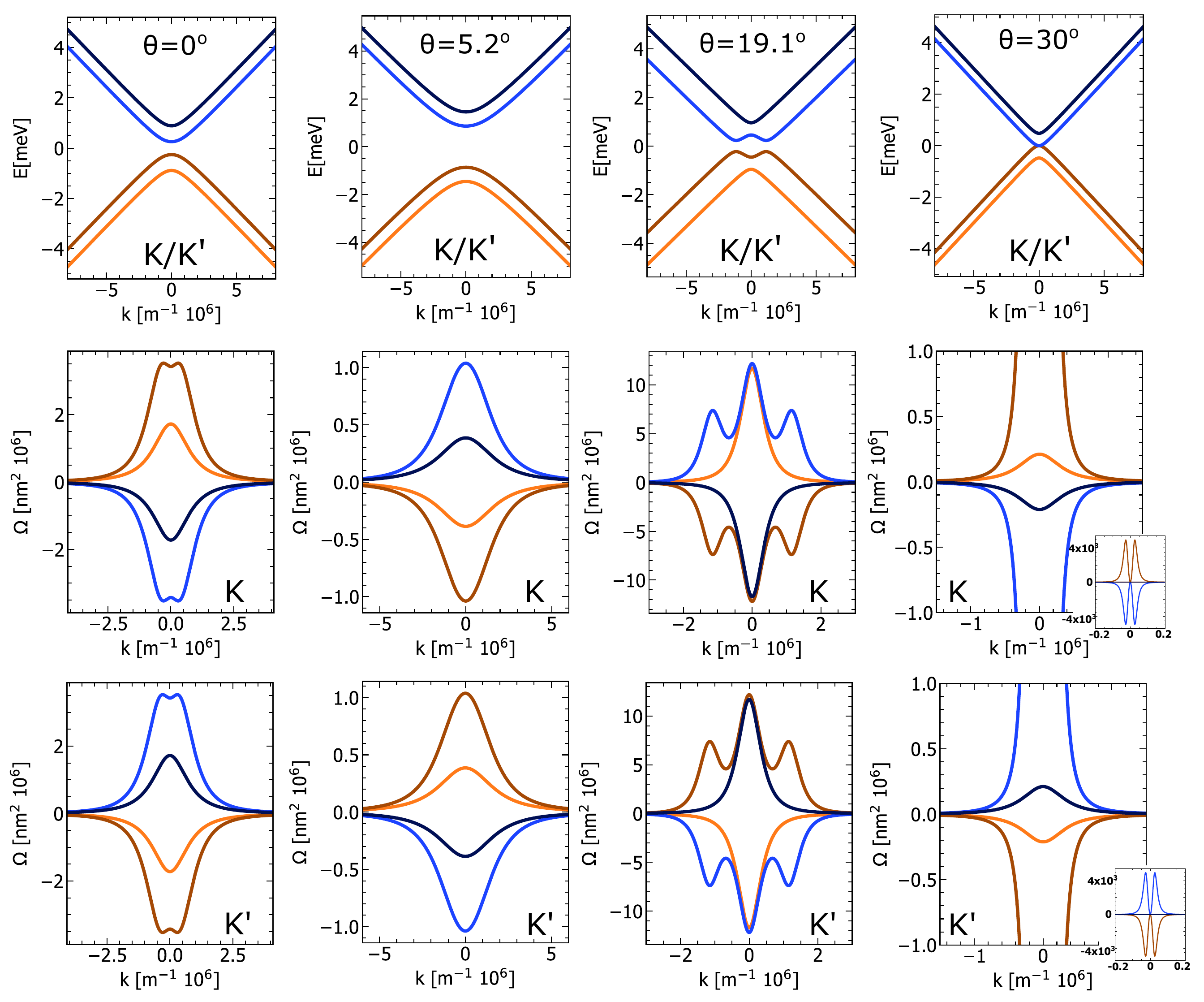}
	\caption{The band structure of t-Gr/MoS$_2$ for the four selected twist angles, $\theta$, and the associated Berry curvatures plotted around the K and K' points, respectively. The colours of individual Berry curvatures correspond to the associated energy band. The values of parameters $\lambda$, $\lambda_{R}$, $\phi_R$ and $\Delta$ for the certain twist angle, $\theta$, are taken from Fig~\ref{fig:Fig1}, the $\upsilon$ parameter is equal  $4.356\cdot10^{-10}$ eVm. }
	\label{fig:Fig5}
\end{figure}

Importantly, the relative twist of graphene with respect to the substrate can substantially change also the orbital and valley dependent phenomena. Here we focus on two important phenomena, that is, on the valley Hall effect and valley Rashba-Edelstein effect. 
The valley Hall effect can be defined as~\cite{Dyrdal2017,wojciechowska_intrinsic_2024,ISLAM2016304}:
\begin{equation}
\sigma_{VH}  = \sigma_{xy}^{s_z\,K} - \sigma_{xy}^{s_z\,K'} 
\end{equation}
where the charge conductivity is give by the general expression:~\cite{Abrikosov,Mahan}
\begin{equation}
\label{eq:sigma_yx}
\sigma_{yx}^{\nu} = \lim_{\omega \to 0} \frac{e^{2} \hbar}{\omega} \int \frac{d \varepsilon}{2\pi} \int \frac{d^{2}\mathbf{k}}{(2\pi)^{2}} \mathrm{Tr} \left[\hat{v}_{y}^{\nu} G^{\nu}_{\mathbf{k}}(\varepsilon + \omega) \hat{\upsilon}_{x}^{\nu}G^{\nu}_{\mathbf{k}}(\varepsilon ) \right],
\end{equation}
%is the transverse component of the conductivity tensor calculated for the $\nu-th$ valley.
Here it should be stressed that the valley Hall conductivity defined above is a special case of the valley orbital Hall effect~\cite{Vignale_OHE_PRB2021}. The consistent theory of the orbital Hall effects is still under development ~\cite{Vignale_OHE_PRB2021,Vignale_PRB2025,Ferreira_PRL2025}, thus for the purpose of this paper we keep the above definition of valley Hall effect, whereas the more detailed discussion of the orbital effects in the graphene based twisted structures will be provided elsewhere.
In the case of intrinsic valley Hall effect the Eq.~(13) leads to the expression connected to  the Berry curvature,$\Omega_{j}$, and has the form:
\begin{equation}
\sigma_{xy}^{\nu} = \frac{e^{2}}{\hbar} \sum_j \int \frac{d^{2} \mathbf{k}}{(2\pi)^{2}} \Omega_{j}^{\nu} f(E_{j})
\end{equation}
where $f(E_{j})$ denotes the Fermi-Dirac distribution function for the $j$-th subband, and $\Omega_{j}^{\nu}$ is the Berry curvature calculated for the $j$-th subband at $\nu$-th valley, i.e., the total Chern number for the specific band is: $\Omega_{j} = \Omega_{j}^{K} + \Omega_{j}^{K'}$. Accordingly, it is clearly seen that valley Hall effect can appear even in the case of vanishing Berry curvature, providing that it is nonzero locally at distinct valleys. This case occurs in considered heterostructures. In Fig.~\ref{fig:Fig5} the angle-integrated Berry curvature is plotted for graphene deposited on MoS$_2$ in the vicinity of the K and K' points and for certain twisted  angles in the range between 0 and 30 degs. The twist angle substantially modulate the local Berry curvature, however the contributions from K and K' points for individual bands are opposite and cancel each other. Accordingly for the Fermi energy located in the band gap, the total Berry curvature is zero, thus also the total Chern number for the fully occupied valence bands is zero. As a result, the anomalous Hall effect does not occur, as should be in systems with the time reversal symmetry, but one can observe, for the Fermi level inside the energy gap, finite and quantized valley Hall conductivity.  
\begin{figure} [t!]
\includegraphics[width=0.7\textwidth]{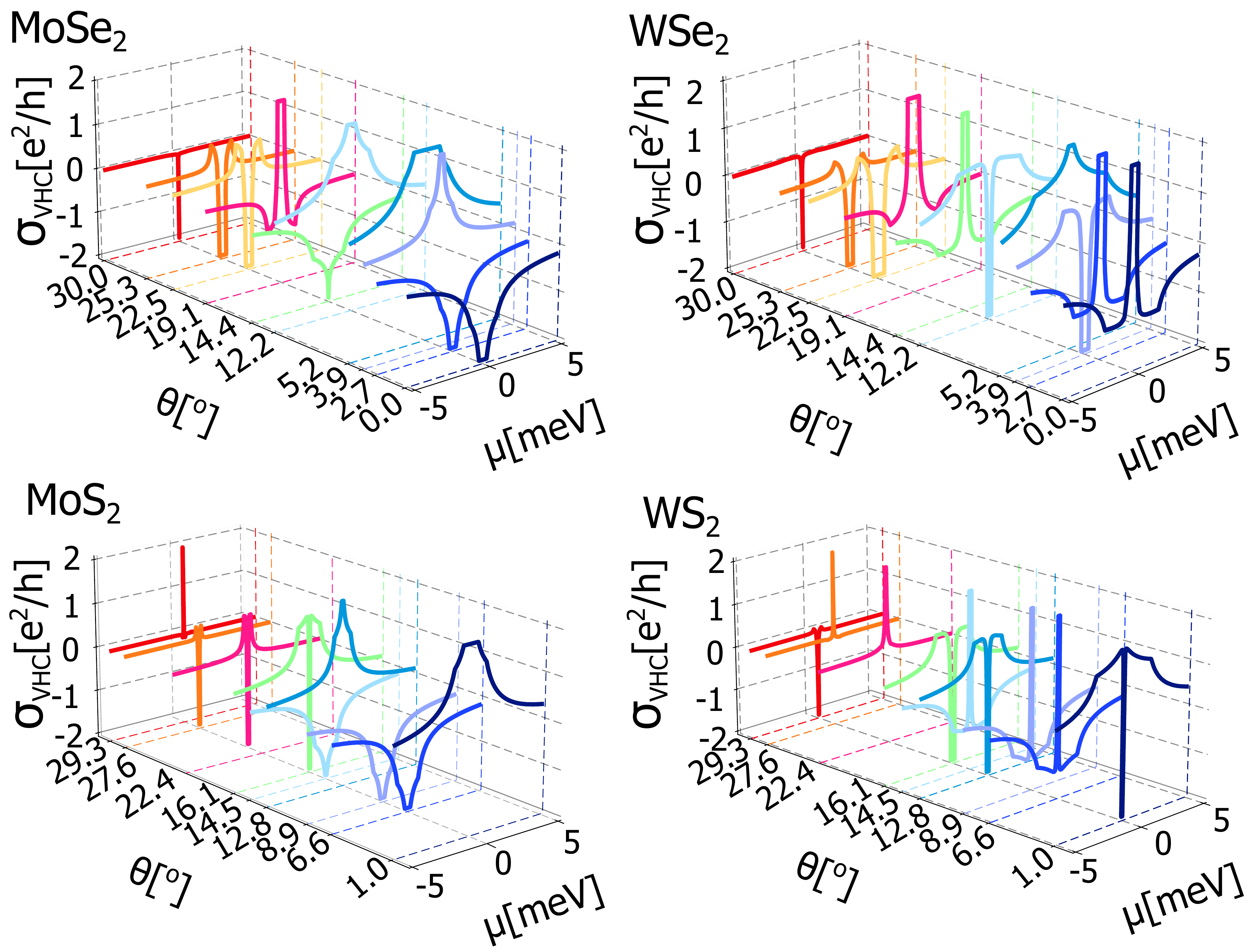}
\centering
	\caption{Valley Hall conductivity of graphene deposited on considered semiconducting TMDC monolayer as a function of the Fermi energy, $\mu$, for certain twist angle between graphene and TMDC, $\theta$. The values of parameters $\lambda$, $\lambda_{R}$, $\phi_R$ and $\Delta$ for the certain twist angle, $\theta$, are taken from Fig~\ref{fig:Fig1}, the $\upsilon$ parameter is equal $5.414\cdot10^{-10}$ eVm for  WSe$_2$, $4.348\cdot10^{-10}$ eVm for WS$_2$, $5.413\cdot10^{-10}$ eVm for MoSe$_2$, and   $4.356\cdot10^{-10}$ eVm for MoS$_2$.}
	\label{fig:Fig6}
\end{figure}
Fig.~\ref{fig:Fig6} present the valley Hall conductivity as a function of Fermi energy and for specific twist angles for all four heterostructures considered in this paper. All figures have been ploted based on analytical formulas obtained based on Eq.~(\ref{eq:sigma_yx}) however, as they are rather long and awkward we decided not to show them. The valley Hall conductivity is symmetric function with respect of change of the sign of Fermi energy and reveals well defined plateau when the Fermi level is located inside the energy gap. Importantly, the valley Hall conductivity takes either +2 or -2 conductance quanta depending on the choice of the twist angle.

\begin{figure*} [t!]
\includegraphics[width=0.75\columnwidth]{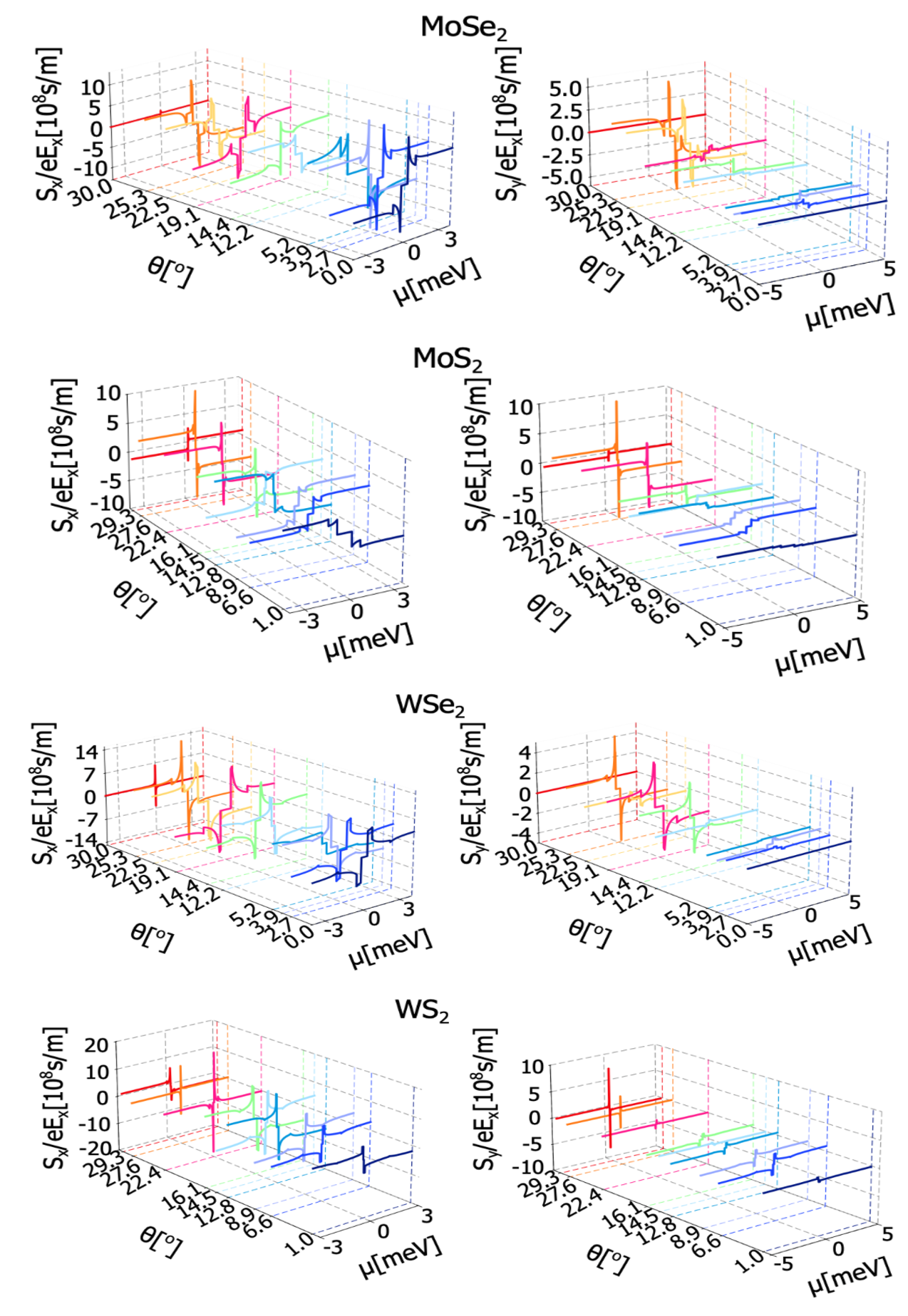}
\centering
	\caption{The \textit{x} and \textit{y} components of nonequilibrium valley spin polarization plotted as a function of Fermi energy, $\mu$, for specific values of the twist angle, $\theta$. The values of parameters $\lambda$, $\lambda_{R}$, $\phi_R$ and $\Delta$ for the certain twist angle, $\theta$, are taken from Fig~\ref{fig:Fig1}, the $\upsilon$ parameter is equal $5.414\cdot10^{-10}$ eVm for  WSe$_2$, $4.348\cdot10^{-10}$ eVm for WS$_2$, $5.413\cdot10^{-10}$ eVm for MoSe$_2$, and   $4.356\cdot10^{-10}$ eVm for MoS$_2$. }
	\label{fig:Fig7}
\end{figure*}

Finally, we have considered the valley Rashba Edelstein eeffect that is a valley nonequilibrium spin polarization that can be defined as~\cite{ISLAM2016304,Manchon_PRB2016,Dyrdal2017,wojciechowska_intrinsic_2024}:
\begin{equation}
S_{\alpha}^{V} = S_{\alpha}^{K}-S_{\alpha}^{K'},
\end{equation}
where $S_{\alpha}^{\nu}$ can be found based on Eq.~(\ref{eq:S_alpha^tau(EF)}).  It has a meaning of emerging nonequilibrium spin imbalance between contribution of the graphene quasiparticles assigned to the two distinct valleys. Fig~\ref{fig:Fig7} presents the valley Rashba Edelstein effect plotted  as a function of Fermi energy  for certain twist angles. The valley spin polarization, similarly as the total spin polarization, is also aligned in plain of the heterostructure and is proportional to the external electric field. It strongly depends not only on the twist angle but also on the Rashba angle, as it is sensitive to the the spin-momentum locking in the structure. Importantly Valley spin conductivity does not depend on the relaxation time, thus it is roboust to the effects of impurities and other disorder responsible for the relaxation proces in the heterostructure.

\section{Discussion and summary}
\label{sec:Sect5}

We have considered graphene deposited on four different semiconducting TMDC monolayers, i.e., on MoSe$_2$, WSe$_2$, MoS$_2$, and WS$_2$.  The main objective was to check how the relative twist angle between graphene and semiconducting TMDC monolayer changes spin-orbital proximity effects and orbital physics in the electronic states of graphene. Accordingly, using effective low-energy Hamiltonian and Green function formalism we calculated the spin Hall effect and nonequilibrium spin polarization in the system. In both cases we have derived analytical formulas, which clearly show that the twist angle can strongly modulate both spin Hall  and spin-to-charge conversion characteristics. One can conclude that the spin Hall effect and nonequilibrium spin polarization in the considered heterostructures  originate from the  proximity-induced Rashba and valley-Zeeman spin-orbit couplings. In consequence, spin is not a good quantum number and therefore spin Hall conductivity contains only the contribution from electrons at the Fermi level. Thus, there is no topological contribution that could lead to the quantum spin Hall insulator phase, when the Fermi energy is located in the energy gap. As the staggered potential and intrinsic spin-orbit coupling are significantly modulated by the twist angle, $\phi$, both the spin Hall conductivity and Rashba-Edelstein effect are significantly modulated by the twist. Importantly, the spin Hall effect does not depend on the Rashba angle, $\phi$, whereas the Rashba-Edelstein effect is substantially affected by the angle $\phi$, what results in an additional component to the spin polarization which is parallel to the external electric field. 
We have also determined the valley Hall effect and valley Rashba-Edelstein effects. Of  course both valley transport phenomena are strongly modulated by the relative twist between graphene and TMDC monolayers. This is natural consequence of the fact that the twist angle strongly modulates the orbital-dependent characteristics and also the local contributions to the Berry curvature. Importantly we have shown that the valley Hall conductivity can take the quantized value equal to $\pm 2 e^2/h$.

It should be noted that in our calculations we have assumed low concentration of short-range, point like nonmagnetic impurities. In other words, the calculations have been done in the constant relaxation time approximation, i.e., the relaxation time, $\tau$, is treated as a constant parameter, and we have not included the impurity vertex correction. It is well known that the impurity vertex correction has a remarkable impact on the spin Hall conductivity in semiconducting heterostructures with Rashba interaction~\cite{Inoue_PRB2004}. However, the vertex correction has a lesser impact on transport characteristics in graphene-based heterostructures, leading to a renormalization of the results obtained in the single-loop approximation by a
factor of the order of unity \cite{Sinitsyn_PRL2006,Dyrdal_PRB2015}.  The evaluation of the impact of disorders on relaxation time and transport characteristic in graphene-based twisted structures is important but extensive issue. Some important results dedicated to this issue have been reported  recently~\cite{Ferreira_PRB2022,Ferreira_PRL2024}. However the effect of disorder in the considered structures needs extended evaluation that treates on the equal footing the impurities as well as fluctuating in space Rashba and intrinsic spin orbit interactions. Accordingly, we decided to restrict considerations presented in this manuscript  only to the clean limit. The more consistent theory describing the effect of the twist angle on the relaxation processes and vertex correction will be presented elsewhere.

\bibliography{bib}

%\noindent LaTeX formats citations and references automatically using the bibliography records in your .bib file, which you can edit via the project menu. Use the cite command for an inline citation, e.g.  \cite{Hao:gidmaps:2014}.

%For data citations of datasets uploaded to e.g. \emph{figshare}, please use the \verb|howpublished| option in the bib entry to specify the platform and the link, as in the \verb|Hao:gidmaps:2014| example in the sample bibliography file.

\section*{Acknowledgements}

This work has been supported by the Norwegian Financial Mechanism under the Polish-Norwegian Research Project NCN GRIEG ’2Dtronics’, project no. 2019/34/H/ST3/00515.

\section*{Author contributions statement}

I.W. performed analytical and numerical calculations, analysed results and wrote the first version of the manuscript. A.D. initiated and supervised the study, analysed results and wrote the manuscript. All authors reviewed the manuscript.

\section*{Additional information}

%To include, in this order: 

\subsection*{Competing interests} 
The authors declare no competing interests.

\subsection*{Correspondence} 
Correspondence and requests for materials should be addressed to I.W. and/or A.D.

%The corresponding author is responsible for submitting a \href{http://www.nature.com/srep/policies/index.html#competing}{competing interests statement} on behalf of all authors of the paper. This statement must be included in the submitted article file.

\end{document}